\documentclass[superscriptaddress,twocolumn,showpacs,pre]{revtex4-1}

\usepackage{graphicx}  
\usepackage{amssymb,amsfonts,amsmath}
\usepackage{color}
\usepackage{epsfig}
\usepackage{epstopdf}
\usepackage{multirow}
\usepackage{natbib}
\usepackage[version=4]{mhchem}

\newcommand{\eps}{\epsilon}
\newcommand{\bra}[1]{\left(#1\right)}

\begin{document}

\title{Frequency locking in auditory hair cells: Distinguishing between additive and parametric forcing}

\author{Yuval Edri}
\affiliation{Department of Physics, Ben-Gurion University of the Negev, Beer-Sheva, Israel}
\affiliation{Physics Department, Nuclear Research Center Negev, P.O. Box 9001, Beer-Sheva 84190, Israel}
	
\author{Dolores Bozovic}
\affiliation{Department of Physics and Astronomy and California NanoSystems Institute, University of California Los Angeles, CA, 90025, US}

\author{Arik Yochelis}
\affiliation{Department of Solar Energy and Environmental Physics, Swiss Institute for Dryland Environmental and Energy Research, Blaustein Institutes for Desert Research, Ben-Gurion University of the Negev, Sede Boqer Campus, 8499000 Midreshet Ben-Gurion, Israel}
\email{yochelis@bgu.ac.il}

\begin{abstract}
The auditory system displays remarkable sensitivity and frequency discrimination, attributes shown to rely on an amplification process that involves a mechanical as well as a biochemical response. Models that display proximity to an oscillatory onset (a.k.a. Hopf bifurcation) exhibit a resonant response to distinct frequencies of incoming sound, and can explain many features of the amplification phenomenology. To understand the dynamics of this resonance, frequency locking is examined in a system near the Hopf bifurcation and subject to two types of driving forces: additive and parametric. Derivation of a universal amplitude equation that contains both forcing terms enables a study of their relative impact on the hair cell response. In the parametric case, although the resonant solutions are 1:1 frequency locked, they show the coexistence of solutions obeying a phase shift of $\pi$, a feature typical of the 2:1 resonance. Different characteristics are predicted for the transition from unlocked to locked solutions, leading to smooth or abrupt dynamics in response to different types of forcing. The theoretical framework provides a more realistic model of the auditory system, which incorporates a direct modulation of the internal control parameter by an applied drive. The results presented here can be generalized to many other media, including Faraday waves, chemical reactions, and elastically driven cardiomyocytes, which are known to exhibit resonant behavior.
\end{abstract}


\maketitle
\newpage

\section{Introduction}

{The sense of hearing requires exquisite mechanical detection, with barely audible tones evoking displacements of the basilar membrane on the order of angstroms~\cite{Hudspeth2014,Hudspeth2005,LeMasurierGillespie2005}.} The auditory system is also highly tuned, with frequency selectivity in various species reaching 0.1\%, and the frequency range reaching as high as 100 kHz. Over the last 60 years, there have been significant advances in our understanding of the inner ear. However, the detailed mechanisms of the auditory system are still not understood, and thus, deficits are being mostly aided by technological solutions, such as cochlear implants.

Nonlinear effects have been shown to be important for the extreme sensitivity and robustness of the inner ear~\cite{RoblesRuggero2001,AshmoreKolston1994,FettiplaceFuchs1999,GillespieDumontKachar2005}. Compressive nonlinearity plays a role both in protecting the cells from damage, and for ensuring that the lowest levels of incoming sound receive the highest degree of amplification~\cite{Hudspeth2014}. Nonlinear response has been demonstrated both at the organism level~\cite{RoblesRuggero2001} and in the motility of individual hair cells~\cite{LeMasurierGillespie2005}. Here, we focus on the latter, since hair cells constitute the main functional elements in the detection process. 

On the apical surface of the hair cell, 20-300 stereocilia comprise the hair bundle; tips of neighboring stereocilia are connected by tip links~\cite{KacharParakkalKurcEtAl2000,Pickles1992}. During stimulus, deflections due to incoming sound induce shearing of the stereocilia comprising the hair bundle, increasing the tension on the tip links between them. The links are coupled to mechanically sensitive ion channels, which  open and close in response to the stimulus forces~\cite{VollrathKwanCorey2007}. The resulting influx of ions depolarizes the cell, and thus leads to the release of neurotransmitters. 

When the bundles are deflected by sound waves, they move in a highly viscous medium. An active process has {therefore} been proposed to explain the high acuity of hearing~\cite{Dallos1992,Hudspeth97,Manley2001,AshmoreAvanBrownellEtAl2010}. Several models {were} developed to explain how the hair cell generates the forces needed to pump energy into the oscillation. These include amplification by active hair bundle motility~\cite{MartinHudspeth1999, VilfanDuke2003,KennedyEvansCrawfordEtAl2003,StaufferHolt2007, BenserMarquisHudspeth1996,ChanHudspeth2005, BeurgNamCrawfordEtAl2008} and electromotility, a process of elongation and contraction of the hair cell soma in response to electrical stimulation~\cite{DallosEvans1995,GeleocHolt2003,DallosWuCheathamEtAl2008,ZhengShenHeEtAl2000,RussellNilsen1997, FisherNinReichenbachEtAl2012, NobiliMammanoAshmore1998}. 

{Empirical evidence provided by otoacoustic emissions \cite{Manley97, Manley2001} indicates an underlying active mechanism in the auditory response. Synchronization of hair bundle oscillations by periodic perturbations, exhibited over a wide range of frequencies~\cite{MartinHudspeth1999, Martin01, Fredrickson-HemsingBozovic2012a}, suggests that the auditory system is analogous to general forced oscillatory media, similar to autocatalysis in chemical media or enzymatic dynamics in physiology~\cite{NicolisPortnow1973,Franck1978,KeenerSneyd1998,Murray2002}. To explain the role of active amplification in hearing, theoretical models proposed that hair cell response follows temporal dynamics that arise through a \textit{Hopf} bifurcation~\cite{ChoeMagnascoHudspeth1998,JulicherProst1997,CamaletDukeJulicherEtAl2000,EguiluzOspeckChoeEtAl2000, DukeJulicher2003,Magnasco2003,KernStoop2003}, a generic mechanism that describes the birth of oscillatory behavior once the critical value of a control parameter is exceeded~\cite{CrossHohenberg1993}. The model equations predict high gain and sharp frequency selectivity at low-amplitude stimuli, and a reduction of both with increasing amplitudes. {The amplification gain was shown to diverge as the control parameter approaches a critical value and to diminish away from the critical point ~\cite{ChoeMagnascoHudspeth1998,EguiluzOspeckChoeEtAl2000,gomez2016signal}.} 
	Consequently, proximity to a Hopf bifurcation is recognized to provide important advantages in explaining the phenomenology of hearing~\cite{Hudspeth2014,HudspethJulicherMartin2010,SzalaiChampneysHomerEtAl2013}.}

{Experimental studies have shown that the dynamic response of a hair cell intertwines many degrees of freedom and complex processes, including biochemical feedback on the control parameter ~\cite{ShlomovitzFredrickson-HemsingKaoEtAl2013,Kao2013,gomez2016signal}. Yet, theoretical models have typically included only additive forcing terms ~\cite{EguiluzOspeckChoeEtAl2000,Fredrickson-HemsingBozovic2012a,KernStoop2003}. We develop a general theoretical framework that allows a systematic study of the impact of parametric versus additive forcing on the resonant response in the cochlea. The two types of forcing reflect the coupling between the driving force and the original (unforced) system, with the parametric term reflecting a situation in which the periodic forcing directly impacts one or more parameters. For example, the light-sensitive oscillatory Belousov--Zhabotinsky chemical reaction under periodic illumination is a parametrically forced system, since the light affects a chemical reaction~\cite{petrov1997resonant,LinBertramMartinezEtAl2000}. The oscillatory nature of the Belousov--Zhabotinsky chemical reaction is analogous to the spontaneous hair bundle oscillations, while the role of illumination~\cite{Bell2012} is analogous to feedback by calcium ions.} 

{Frequency locking is a generic feature of periodically driven oscillatory systems that exhibit resonant behavior, examples of which include Faraday waves, nonlinear optical solitons, Josephson junctions, and chemical reactions. Frequency locked response has also been shown to be a crucial feature of auditory detection~\cite{Hudspeth2014}. We examine the properties of the 1:1 resonance domain (Arnold Tongue) in the cochlear response. We consider hair cells to be poised in the vicinity of the Hopf bifurcation~\cite{Hudspeth2014} and {derive a universal normal form equation that includes both additive and parametric driving forces}. {Specifically, we examine the distinctions in the transition from unlocked to locked oscillations, under different types of forcing.} This model of the cochlear response allows for the coexistence of multi-modal frequency locking (i.e., beyond 1:1 resonance) and elucidates the presence of super- vs. sub-critical forms of the frequency locking transition~\cite{Hudspeth2014}. Thus, this study provides a framework for incorporating both biochemical and mechanical feedback in the description of the auditory system. }


\section{{Frequency locking under additive vs. parametric forcing}}

{{Periodically forced oscillatory systems can be mathematically represented as follows:}
	\begin{eqnarray}\label{eq:genODE}
	\nonumber	{\dfrac{{{\rm{d}}u_1 }}{{{\rm{d}}t}}}&=& {f_1 \left( {\vec u} \right) + g_1 \left( {\vec u} \right)\cos \omega _f t},  \\ 
	{\dfrac{{{\rm{d}}u_2 }}{{{\rm{d}}t}}}&=& {f_2 \left( {\vec u} \right) + g_2 \left( {\vec u} \right)\cos \omega _f t},  \\
	\nonumber	&\vdots&   \\
	\nonumber	{\dfrac{{{\rm{d}}u_N }}{{{\rm{d}}t}}}&=& {f_N \left( {\vec u} \right) + g_N \left( {\vec u} \right)\cos \omega _f t},
	\end{eqnarray}
	where $\vec{u}=(u_1,u_2,...,u_N)$ is a set of observables (with $N$ being an integer)}, $f$ and $g$ denote functions that describe interactions {between observables}, and $\omega_f$ is the frequency of the driving force. {In what follows, we will consider, without a loss of generality, the two variable activator--inhibitor FitzHugh-Nagumo (FHN) model~\cite{hugh1961impulse,nagumo1962active}, in which $\vec{u}=(u_1,u_2):=(u,v)$. {Forcing acts on the activator variable $u$; up to linear order, it is given by $g_1=\gamma_a+\gamma_p u+{higher~~order~~terms}$, and $g_2=0$. {We note that higher order forcing terms do not contribute to the discussed results, which focus on 1:1 resonance}. Magnitudes of the additive and parametric forcing terms are represented by $\gamma_a$ and $\gamma_p$, respectively. Additional details on the FHN model are given in the Appendix. } }

{Near the Hopf bifurcation and in the absence of periodic forcing, model equations of type~\eqref{eq:genODE} can be reduced to a universal nonlinear equation, which is often referred to as the Stuart-Landau (a.k.a. complex Ginzburg-Landau) equation~\cite{CrossHohenberg1993}. Under externally applied periodic forcing (at frequency $\omega_f$), the phase invariance along the limit cycle is destroyed, {and the system exhibits either unlocked oscillations or entrained oscillations with discrete phase shifts}. {Increasing the amplitude of the drive increases the range of detuning under which a resonant solution can arise. This detuning} from an unforced }characteristic Hopf frequency $\omega_c$ is given by $\nu=\omega_c-\omega_f/n$ and thus corresponds to $n:1$ frequency locking (where $n$ is an integer), for which the original system responds at translations $t \to 2\pi/\omega_f$ ~\cite{CoulletEmilsson1992}. Under small detuning values $|\nu| \ll \omega_c$, the Stuart-Landau equation is modified as~\cite{Gambaudo1985,ElphickIoossTirapegui1987}:
\begin{equation}\label{eq:FCGL}
\frac{\mathrm{d} A}{\mathrm{d} t}= \bra{\mu+i\nu}A+\bra{1+i\beta}|A|^2A+\Gamma \bar{A}^{n-1},
\end{equation}
where $A$ is a complex amplitude that describes weak temporal modulations of a primary limit cycle that is generated at the Hopf onset, $\mu$ measures the distance from the Hopf bifurcation, $\beta$ is the nonlinear frequency correction, $\Gamma$ is a (real) forcing magnitude, and $\bar{A}$ is the complex conjugate of $A$. Notably, the $n:1$ resonant solutions are invariant under $A\to Be^{-i2\pi/n}$~\cite{CoulletEmilsson1992}; hence, the frequency locking condition implies that $B$ is constant. This resonance condition is fulfilled over a finite range of driving frequencies and amplitudes, which defines the domain of an Arnold Tongue. 

We focus this study on the 1:1 resonant response~\cite{HudspethJulicherMartin2010,Fredrickson-HemsingBozovic2012a}. The amplitude equation~(\ref{eq:FCGL}) with $n=1$ has been employed in several contexts, such as studies of fluctuations and the response to pitches~\cite{Balakrishnan2005,martignoli2013pitch}. {Here, we examine the physical nature of the forcing and its implications for the resonant response.} In what follows, we show that for 1:1 resonance, one can obtain a generalized amplitude equation 
\begin{equation}\label{eq:modFCGL}
\frac{\mathrm{d} A}{\mathrm{d} t}=\left(\mu+i\nu\right) A -\left(1+i\beta\right)|A|^2A + \Gamma_p\bar{A}+\Gamma_a,
\end{equation}
{and explore its} frequency locking properties; for details, we refer the reader to the Appendix.

\section{Phase locked solutions and 1:1 resonance domains}

{To determine the regions of frequency locked solutions, we rewrite \eqref{eq:modFCGL} and separate the contributions of the additive and parametric components by the transformation}
\begin{equation}\label{add_trans}
A \to A \exp\left(-i\tan^{-1}\frac{\Im m \Gamma_{a,p}}{\Re e\Gamma_{a,p}}\right).
\end{equation}
{Next, we introduce the following notation to distinguish the additive and the parametric forcing terms:}
\begin{equation}\label{eq:forcing_type_param}
\delta=\Bigg\{\begin{array}{cc}
0 & \gamma_a >0,~~\gamma_p=0\\
1 & \gamma_a =0,~~\gamma_p>0
\end{array},
\end{equation}
for which:
\begin{equation}
\nu_\delta := 2\nu-\mu\omega_c+\delta~ \frac{1+\omega_c^2}{(\nu-3\omega_c)(\nu+\omega_c)}\frac{\gamma_p^2}{2\omega_c}.
\end{equation}
Using the polar representation $A=\rho e^{i\phi}$ and looking for stationary solutions, we impose the conditions:
\begin{subequations}\label{eq:amp_rho_phi}
	\begin{eqnarray}
	\mu \rho-\rho^3+\delta|\Gamma_a|\cos\phi+\bra{1-\delta}|\Gamma_p|\rho \cos 2\phi=0,\\
	\nu_\delta -\beta\rho^2-\delta \rho^{-1}|\Gamma_a|\sin\phi-\bra{1-\delta}|\Gamma_p| \sin 2\phi=0.
	\end{eqnarray}
\end{subequations}
Solutions to~(\ref{eq:amp_rho_phi}) thus satisfy an equation for the amplitude~\cite{MaBurkeKnobloch2010}:
\begin{equation}\label{eq:amp_rho}
(\rho^2-\mu)^2+\bra{\beta\rho^2-\nu_\delta}^2=\delta\rho^{-2}|\Gamma_a|^2+\bra{1-\delta}|\Gamma_p|^2,
\end{equation}
and for the phase
\begin{subequations}\label{eq:amp_phi}
	\begin{eqnarray}
	\label{stat_phase_cos}
	\cos{\bra{1+\delta}\phi}&=&\rho\frac{\rho^2-\mu}{\delta|\Gamma_a|+\bra{1-\delta}|\Gamma_p|\rho}, \\
	\sin{\bra{1+\delta}\phi}&=&\rho\frac{\nu_\delta-\beta\rho^2}{\delta|\Gamma_a|+\bra{1-\delta}|\Gamma_p|\rho}.
	\label{stat_phase_sin}
	\end{eqnarray}
\end{subequations}

For the additive case ($\delta=0$), the phase displays $2\pi$ symmetry shifts, while for the parametric case ($\delta=1$), one obtains phase shifts of $\pi$. The latter implies bistability of frequency locked solutions, viz. $A_0=\tilde \rho e^{i \tilde \phi}$ and $A_\pi=\tilde \rho e^{i(\tilde \phi+\pi)}$, where $\tilde \rho$ and $\tilde \phi$ are solutions to~(\ref{eq:amp_rho_phi}). {This result is analogous to a situation where a spatially periodic system is driven by a space-dependent modulation~\cite{MauHaimHagbergEtAl2013}. Thus, the $1:1$ resonant solutions {can exhibit different phase symmetries, with one of them displaying} properties similar to the $2:1$ resonant solutions~\cite{BurkeYochelisKnobloch2008}. The linear stability of these solutions 
	\begin{equation}\label{eq:linear}
	\left( {\begin{array}{*{20}c}
		\rho   \\
		\phi   \\
		\end{array}} \right) - \left( {\begin{array}{*{20}c}
		{\tilde \rho }  \\
		{\tilde \phi }  \\
		\end{array}} \right) \propto e^{\sigma t},
	\end{equation}}
is determined by the sign of the real part of the eigenvalues 
\begin{widetext}
\begin{equation}\label{eq:eigenvalues}
\sigma_{\pm}=\mu-2\rho^2\pm \sqrt{\bra{\mu-2\rho^2}^2-
\left[\bra{\delta+3}\rho^4 \bra{1+\beta^2}+4\rho^2\bra{\mu+\beta\nu_\delta}+\delta\bra{\mu^2+\nu_\delta^2}\right]},
\end{equation}
\end{widetext}
where the solutions are stable if $\Re e\sigma_\pm<0$ and unstable otherwise.

\begin{figure}
	(a)\centerline{\includegraphics[width=.4\textwidth]{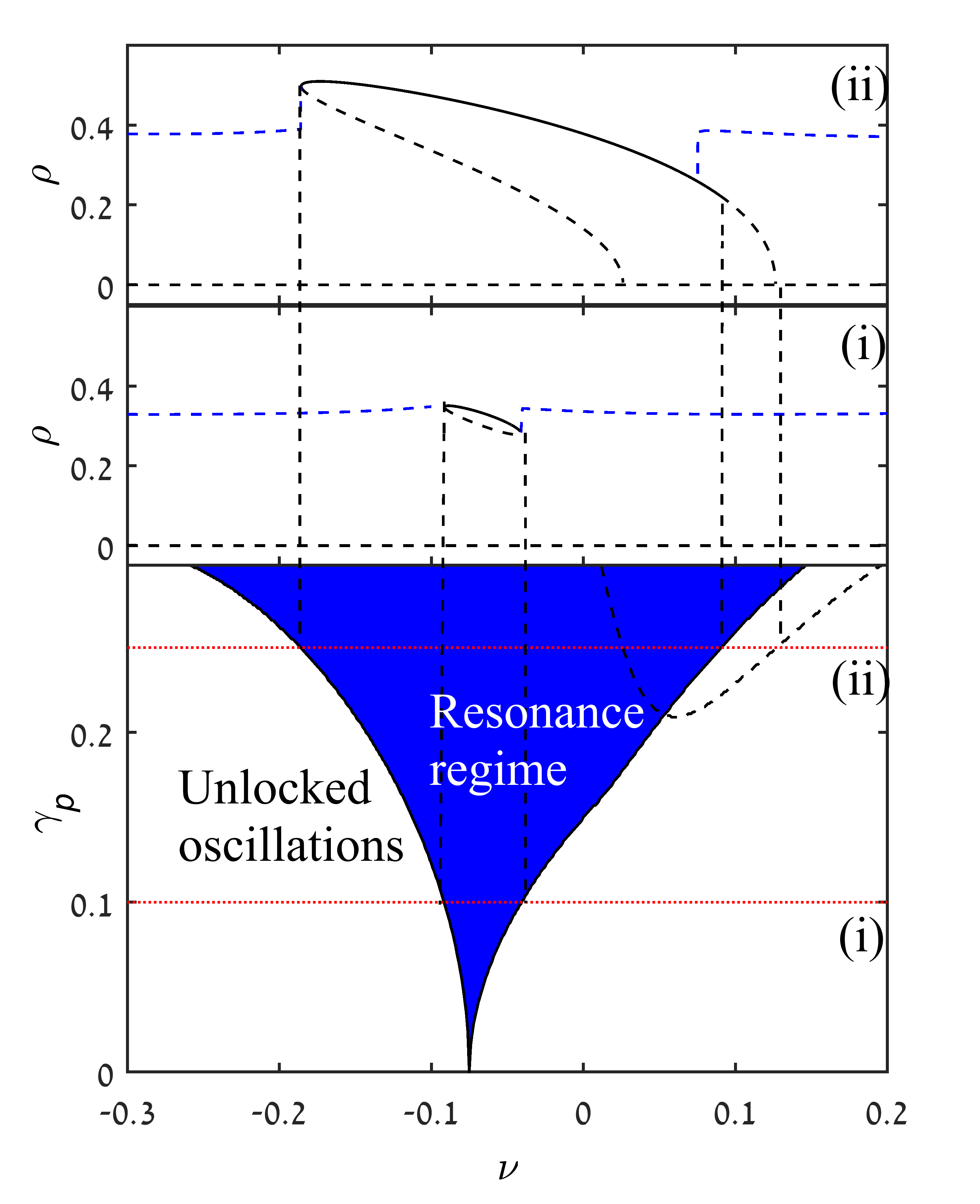}}
	(b)\centerline{\includegraphics[width=.4\textwidth]{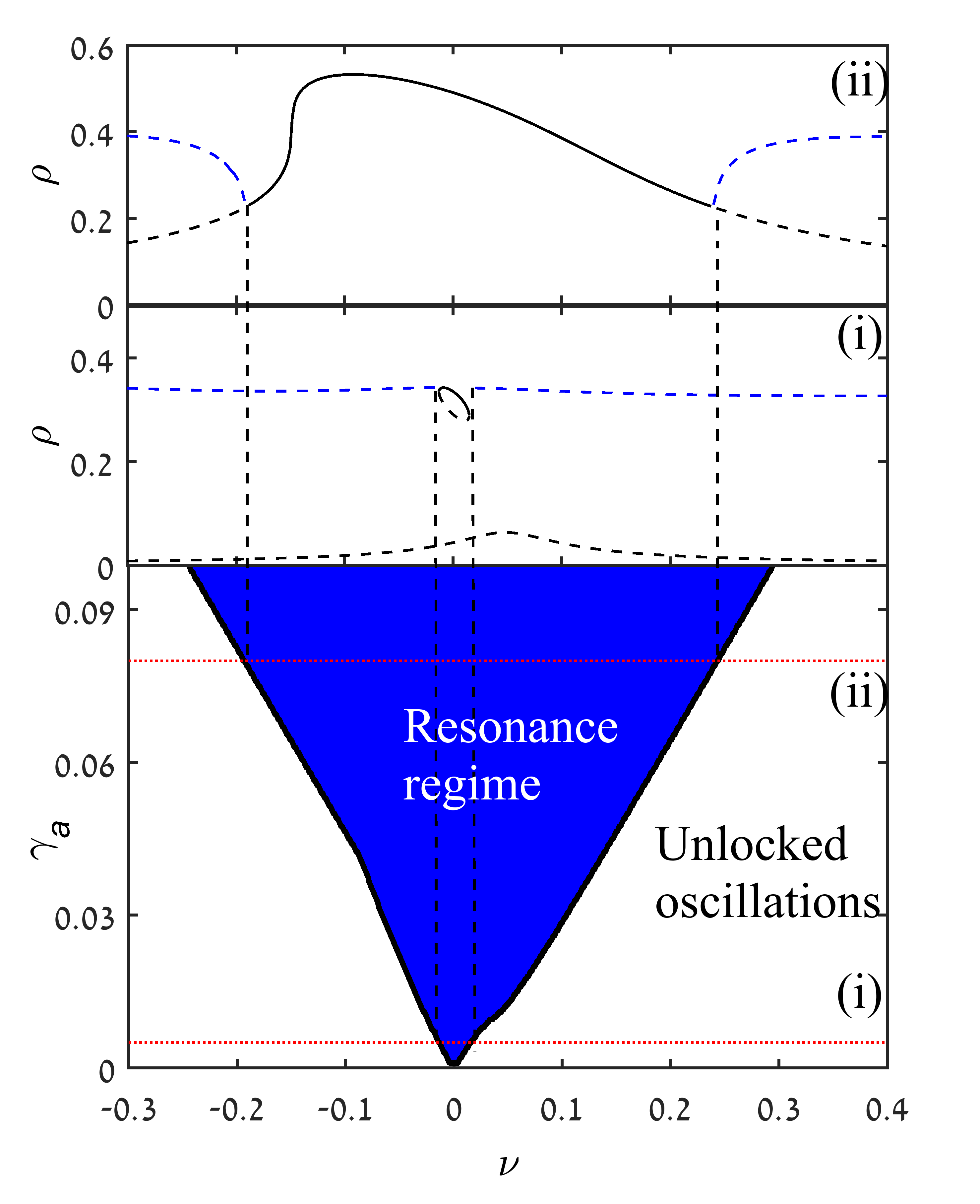}}
	\caption{(color online) Domains describing the 1:1 frequency locked (resonant) and unlocked oscillations above the Hopf bifurcation ($\mu>0$), in a parameter space of detuning ($\nu$) and forcing magnitudes for (a) parametric forcing, $\gamma_p>0, \gamma_a=0$ and (b) additive forcing, $\gamma_p=0, \gamma_a>0$. The bottom panel describes the resonant region (shaded area), while the top panels describe the amplitudes of resonant solutions and unlocked oscillations (light dashed lines) at two distinct $\gamma$ values ($\gamma_p=0.1,0.25$, $\gamma_a=0.005,0.08$) as a function of $\nu$; solid lines in top/mid panels mark stable solutions, and the dashed line in the bottom panel of (a) marks the locus of points at which the nontrivial solutions bifurcate. Equation~\ref{eq:modFCGL} was solved with parameters: $\mu=0.1$ and (a) $\omega_c=0.5$, (b) $\omega_c=1$.}
	\label{fig:above_Hopf}
\end{figure}

\begin{figure}
	(a)\centerline{\includegraphics[width=.4\textwidth]{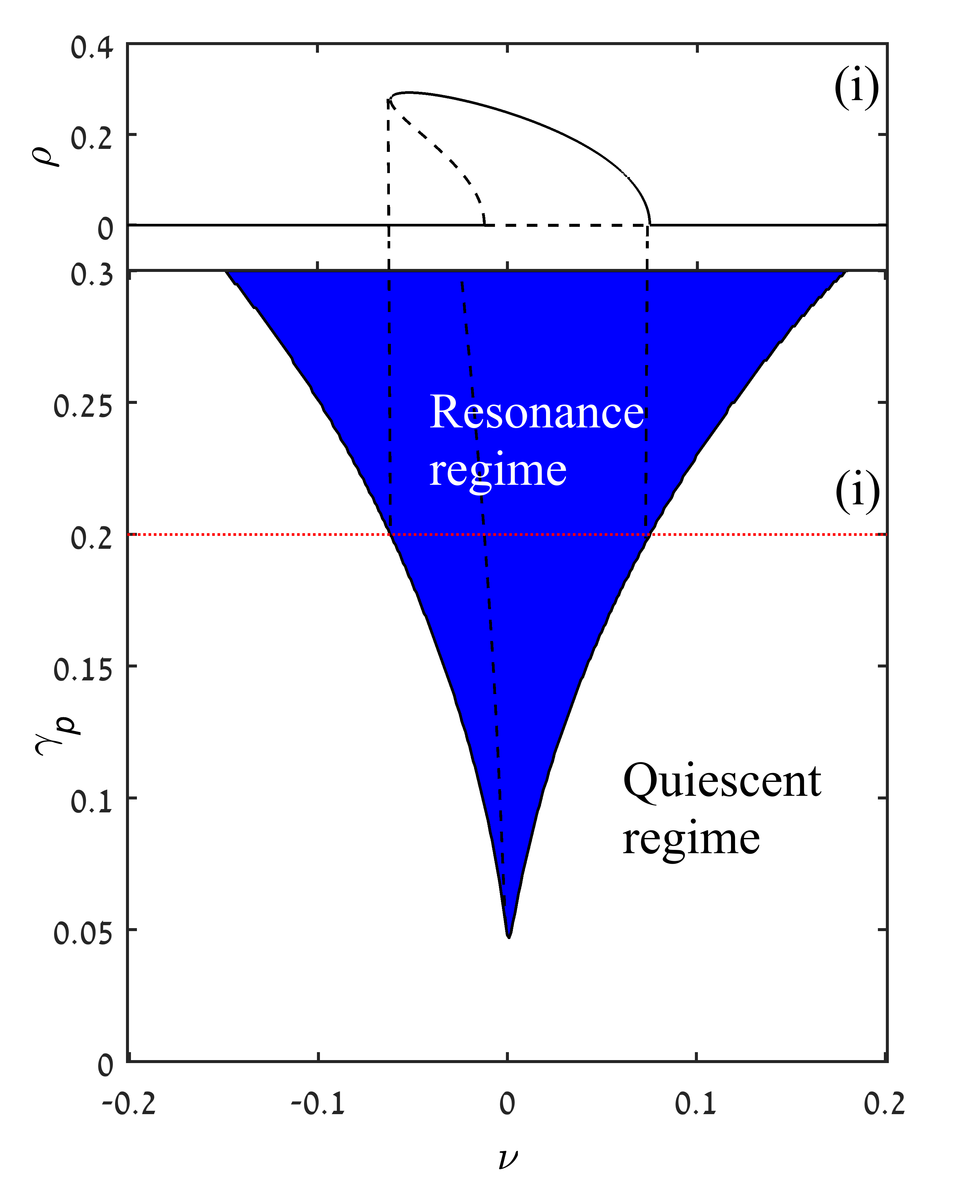}}	
	(b)\centerline{\includegraphics[width=.39\textwidth]{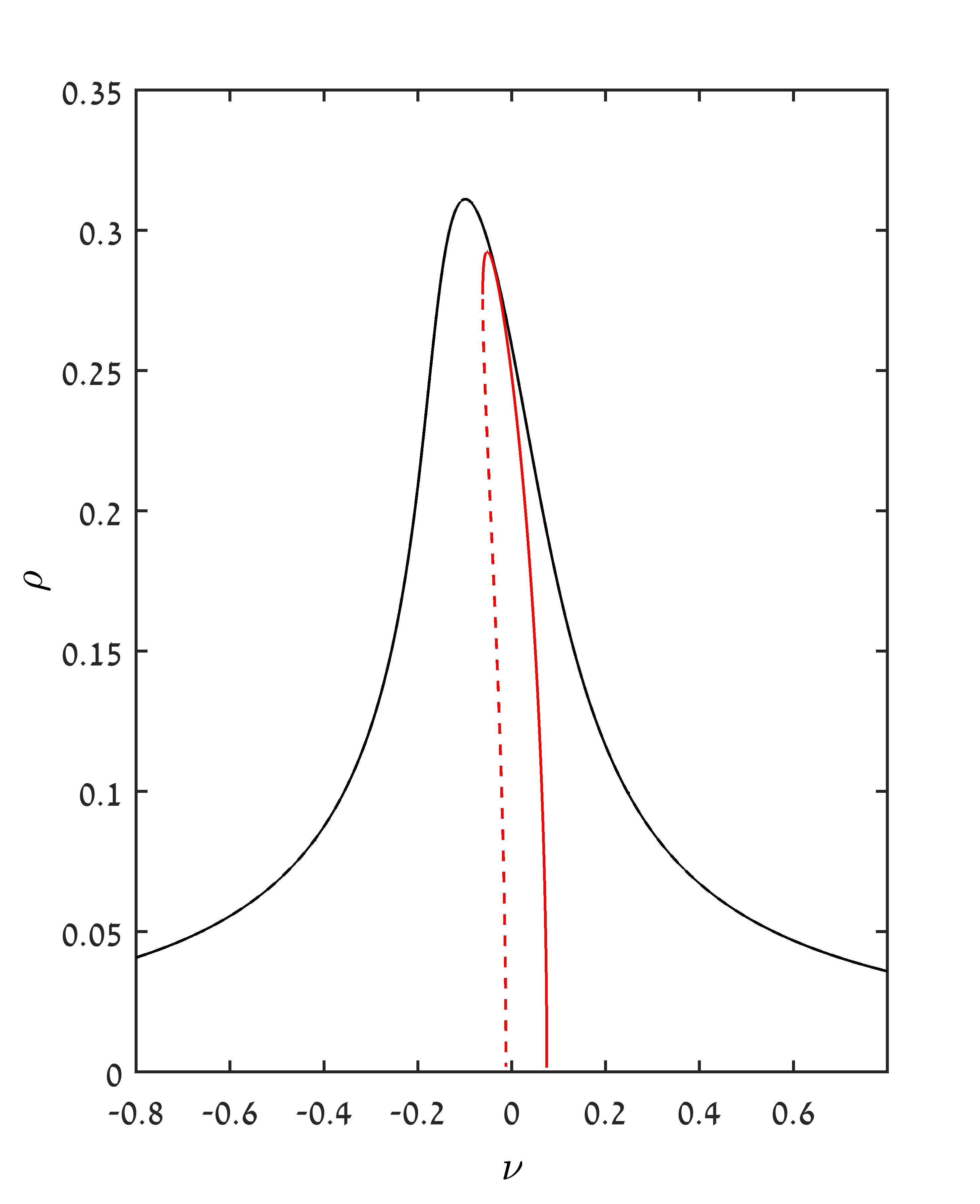}}
	\caption{(color online) (a) 1:1 frequency locked (resonant) domain and unlocked oscillations below the Hopf bifurcation ($\mu<0$), in a parameter space of detuning ($\nu$) and forcing magnitudes for parametric forcing, $\gamma_p>0, \gamma_a=0$. The bottom panel describes the resonant region (shaded area), while the top panel describes a typical behavior at a specific $\gamma$ ($\gamma_p=0.2$) value as a function of $\nu$; solid lines in the top panel mark stable solutions, and the dashed line in the bottom panel marks the locus of sub-critical bifurcation onsets for the nontrivial solution. (b) A typical amplitude dependence at a specific $\gamma$ value ($\gamma_a=0.05$) as a function of $\nu$, for the additive case {(dark line), with a superimposed parametric case (light line) that is taken from (a) at slice (i)}. Equation~\ref{eq:modFCGL} was solved with parameters: (a) $\mu=-0.005$, $\omega_c=0.5$ and (b) $\mu=-0.1$, $\omega_c=1$.}\label{fig:below_Hopf}
\end{figure}

{Combining the {results on existence and stability of frequency locked solutions, we obtain the resonance regimes, for parameter space that} is spanned by the forcing amplitude ($\gamma_{a}$ or $\gamma_{p}$) and detuning ($\nu$).} In particular, we distinguish between two cases: the oscillatory regime ($\mu>0$) and the {quiescent} regime ($\mu<0$), as shown in Fig.~\ref{fig:above_Hopf} and Fig.~\ref{fig:below_Hopf}, respectively.


\subsection{{Frequency locking in the spontaneously oscillating regime}}
Above the onset of the Hopf bifurcation, both the parametric and the additive forcing terms lead to a similar Arnold Tongue response, as shown by the shaded domains in Fig.\ref{fig:above_Hopf}. At weak forcing magnitudes ($\gamma_{a,p} \ll 1$), the frequency locked solutions {form isolas} and coexist with an additional unstable solution; in the parametric case, this unstable state is a trivial one. The shaded region is identified with solutions that are linearly stable {with respect to} temporal perturbations. Outside of the resonance domain, unlocked oscillations prevail, with the maximal amplitude denoted by the light dashed line, as shown in the {horizontal slices (i) at fixed} $\gamma$.

At larger values of $\gamma_p$, two new solutions bifurcate from the trivial state (see the dark dashed lines in the resonance domain of Fig.~\ref{fig:above_Hopf}(a)), where the right line denotes solutions of a super-critical nature, while the left is sub-critical; both are characteristic of the 2:1 resonance~\cite{BurkeYochelisKnobloch2008}. The left non-trivial solution continues towards negative detuning values, folds back toward positive values, and connects with the right super-critical branch. {Stability of this top branch defines the resonant solutions. In the additive forcing case ($\gamma_a$), the isola merges with the bottom solution through a cusp bifurcation~\cite{SzalaiChampneysHomerEtAl2013}, which results in a single amplitude throughout the whole detuning region (not shown here). Both behaviors are described by slices (ii) at fixed $\gamma$. {Further details on the coexistence and stability of such solutions are of secondary significance and will be discussed elsewhere.} }

\subsection{{Frequency locking in the quiescent regime}}
Properties of the frequency locked solutions in the {quiescent} regime are fundamentally different from those of the {innately} oscillatory regime. In the parametric case, the resonant solutions also take the form of an Arnold Tongue. However, outside of the resonance region, the trivial state is stable, and thus the system can be either quiescent (outside of the resonance regime) or frequency locked (within the resonance regime). Nevertheless, since the resonance boundaries preserve the super- and sub-critical properties, part of the resonance boundary exhibits \textit{hysteresis}, as shown in Fig.~\ref{fig:below_Hopf}(a). These properties imply a smooth transition to {{frequency locked oscillations} if one approaches the Arnold Tongue from large positive detuning, and an abrupt transition if the resonance is approached from negative detuning values. The former case is reversible upon detuning, while the latter is associated with two distinct transition onsets.
	
	The frequency locked solutions in the additive case persist throughout the whole parameter range, as shown in Fig.~\ref{fig:below_Hopf}(b) by a typical slice along $\gamma$. Nevertheless, even in the absence of a distinct {transition to resonance, we can identify amplification of the response amplitude.} Notably, while the resonant behavior is smooth for the FHN model, it is possible to observe a hysteresis here as well, via a cusp bifurcation~\cite{MaBurkeKnobloch2010}. However, the hysteresis designates a transition from one oscillatory state to another, rather than a transition from the quiescent state, as in the parametric case.
	
	\section{Discussion}
	
	{ Hair cells were shown to be the sources of amplification in the inner ear ~\cite{LeMasurierGillespie2005}, operating either via hair bundle motility~\cite{KennedyEvansCrawfordEtAl2003,ChanHudspeth2005,BeurgNamCrawfordEtAl2008} or somatic electromotility~\cite{DallosEvans1995,DallosWuCheathamEtAl2008,ZhengShenHeEtAl2000,NobiliMammanoAshmore1998}.  According to theoretical models, the amplification gain and frequency selectivity of the auditory response are dependent on the value of an internal control parameter. As empirical evidences support the existence of a feedback mechanism, which modulates this control parameter in response to external forcing \cite{Kao2013}, it is important to incorporate its dynamics into the theoretical models. A universal framework has therefore been formulated to describe the 1:1 frequency locking of a system poised near the Hopf bifurcation and exposed to both additive and parametric forcing. While additive forcing ($\gamma_a$ term in~(\ref{eq:modFCGL})) has been employed in previous studies, we incorporate here a periodic forcing term that couples directly to biochemical processes ($\gamma_p \bar{A}$ term in~(\ref{eq:modFCGL})) modulating the internal control parameter. {A related study of synchronization dynamics of coupled oscillators also supports the inclusion of biochemical feedback~\cite{gomez2016signal}. We discuss below the empirical data indicating the influence of both types of forcing on the frequency locked response.}
		
		\subsection{Response vs. forcing magnitude}
		{
			{For the 1:1 resonance, we explored the scaling of the response amplitude with the additive and parametric forcing magnitudes.} As a result, we obtain responses that are characterized by distinct power laws. {A number of power laws have been experimentally observed in the response of the auditory system and are discussed in~\cite{MartinH01}.} The experimental results show three distinct regimes in the response: (i) at low forcing magnitudes, the response scales linearly, (ii) as the magnitude is increased, there is a crossover to a nonlinear regime {with $1/3$ exponent, and (iii) an additional transition is observed at high forcing magnitudes. The additive forcing can explain the first two cases, which result from the competition between nonlinear and linear terms. However, the crossover to linearity in the third regime~\cite{SzalaiChampneysHomerEtAl2013,gomez2016signal} is not captured by including only the additive forcing term. The response to parametric forcing exhibits linear scaling $\gamma_p \sim \rho$, and hence supports the emergence of the third regime.}}
		
		\subsection{Phase shifts in frequency locked solutions}
		Most of the model equations have employed additive forcing, where the 1:1 resonant solutions obey $2 \pi$ symmetry~\cite{CoulletEmilsson1992}. The generalized equations developed here show that the inclusion of a parametric forcing term introduces bistability of solutions differing by phase shifts of $\pi$. Phase shifts of $\pi$ have apparently been observed in experiments~\cite{Fredrickson-HemsingBozovic2012a,RuggeroNarayanTemchinEtAl2000}, but not accounted for in a theoretical model.  
		
		\subsection{Hysteresis in the resonance boundaries}
		Construction of resonance domains (i.e., Arnold Tongues) provides insight into the transition from unlocked to frequency locked dynamics. Indeed, experimental measurements have been performed and revealed complex dynamics in the transition from spontaneous oscillation into the resonance regime~\cite{Fredrickson-HemsingBozovic2012a, Maoileidigh13, Salvi15}. Experimental results have raised conjectures on the possible presence of both super- and sub-critical forms of the Hopf instability~\cite{Hudspeth2014}. Our results demonstrate that a hysteresis can be described solely by bifurcations of the phase-locked states, which leave the super-criticality of the Hopf bifurcation intact. Resolving the contributions from the additive versus parametric forcing is difficult in the regime above the Hopf onset (spontaneously oscillatory regime), as the magnitude of unlocked oscillations is equal to that of the locked states. However, the differences become clear in the regime at or slightly below the Hopf onset, where the hysteresis is conjectured to occur in the transition from a quiescent to a frequency locked state.
		
		The results presented here are not limited to a specific model and should arise as general features of 1:1 frequency locking~\cite{Hudspeth2014,SzalaiChampneysHomerEtAl2013} in other systems, such as elastically driven cardiomyocytes~\cite{cohen2016elastic}. We believe that this framework provides a more realistic description of the biological system, as it can incorporate both biochemical and mechanical feedback in descriptions of the auditory response. Further, the theoretical framework that incorporates parametric behavior could be generalized to include {the coexistence of multiple $n:1$ resonances}, which have not been included in previous spatially extended models. Hence, the methodology developed here could provide a framework for future spatiotemporal models of the cochlear response. Finally, this study of frequency locking dynamics can be generalized to other systems, such as Faraday waves, shaken granular media, forced oscillatory chemical reactions, and elastically forced cardiomyocytes.}}

\begin{acknowledgements}
	We thank Robijn Bruinsma, Oreste Piro, and Ehud Meron for helpful discussions on the subject. This work was partially supported by the Adelis foundation (A.Y.)‘, and by NIH grant R01DC011380, NSF grant IOS-1257817 (DB).	
\end{acknowledgements}

\section*{Appendix: Weakly non-linear form of the FitzHugh-Nagumo model}

The FitzHugh-Nagumo model is a general Bonhoeffer-van der Pol type equation, which has been employed as a prototypical model for many biological and chemical systems~\cite{Murray2002}. The response of the system to periodic forcing is described by:
\begin{subequations}\label{eq:FHN}
	\begin{eqnarray}
	\frac{\mathrm{d} u}{\mathrm{d} t}&=&u-u^3-v+\left(\gamma_a+\gamma_p u\right)\cos(\omega_f t),\\
	\frac{\mathrm{d} v}{\mathrm{d} t}&=&\epsilon(u-a v),
	\end{eqnarray}
\end{subequations}
where $u$ is an activator, $v$ is an inhibitor, and $\epsilon$ and $a$ are parameters. We note that the Bonhoeffer-van der Pol type equations have already been employed in modeling the dynamics of the auditory system~\cite{GelfandPiroMagnascoEtAl2010,StoopKernGopfertEtAl2006}.
The trivial solution to the unforced equation (\ref{eq:FHN}) crosses the Hopf instability at $\eps=\eps_c=a^{-1}$ and a critical frequency $\omega_c=\sqrt{\epsilon_c-1}$. Near the instability onset, $\mu := (\epsilon_c-\epsilon)/\epsilon_c\ll 1$, and under 1:1 periodic forcing, Eqs.~\ref{eq:FHN} obey the approximation:
\begin{equation}\label{eq:approx1}
\left(\begin{array}{ccc}
u\\
v
\end{array}\right) \approx \mu^{1/2}\left(\begin{array}{ccc}
u_1\\
v_1
\end{array}\right) + \mu \left(\begin{array}{ccc}
u_2\\
v_2
\end{array}\right) + \mathcal{O}\bra{\mu^{3/2}},
\end{equation}
where
\begin{equation}\label{approx_first_order}
\left(\begin{array}{ccc}
u_1\\
v_1
\end{array}\right)=\left(\begin{array}{ccc}
1\\
1-i\omega_c
\end{array}\right)A\bra{\mu t}e^{i\omega_ft}+\mathrm{complex\, \, conjugate}.
\end{equation}
Employing standard multiple time-scale expansion, and letting $\gamma_p \sim \mu^{1/2}$ and $\gamma_a\sim \mu^{3/2}$~\cite{MauHaimHagbergEtAl2013}, we obtain, up to order $\mu^{3/2}$, the generalized amplitude equation for 1:1 forcing that incorporates both additive and parametric components:
\begin{equation}\label{eq:modFCGL_ap}
\frac{\mathrm{d} A}{\mathrm{d} \tau}=\left(\mu+i\nu_\delta \right) A -\left(1+i\beta\right)|A|^2A + \Gamma_p\bar{A}+\Gamma_a,
\end{equation}
where,
\begin{eqnarray}
\nonumber &&\beta=-\omega_c^{-1},~ \nu_\delta=2\nu-\mu \omega_c+\delta \frac{1+\omega_c^2}{(\nu-3\omega_c)(\nu+\omega_c)} \frac{\gamma_p^2}{2\omega_c}, \\
\nonumber &&\nu=\omega_c-\omega_f, \tau=t/2,~ \Gamma_a=\bra{\frac{\sqrt{3}}{2}-i\frac{\sqrt{3}}{2\omega_c}}\gamma_a, \\
\nonumber &&\Gamma_p=\left(\frac{1}{\omega_c+\nu}-i\frac{1+\omega_c\nu}{\omega_c^2-\nu^2}\right)\frac{\gamma_p^2}{4\omega_c}.
\end{eqnarray}

%

\end{document}